\documentclass[traditabstract]{aa} 
                                   
\usepackage{graphicx}
\usepackage{natbib}
\usepackage{txfonts}

\begin{document}

\title{VLT/X-shooter observations and the chemical composition\\ of cool white 
dwarfs
\thanks{Based on observations collected at the European Organisation for 
Astronomical Research in the Southern Hemisphere, Chile under programme 
ID 080.D-0521, 082.D-0750, 083.D-0540, 084.D-0862 and 086.D-0562.}}

\author{Ad\'ela Kawka\inst{1,2} \and St\'ephane Vennes\inst{1,2}}

\institute{Astronomick\'y \'ustav, Akademie v\v{e}d \v{C}esk\'e republiky, Fri\v{c}ova 298, CZ-251 65 Ond\v{r}ejov, Czech Republic\\
\email{kawka,vennes@sunstel.asu.cas.cz} 
\and
Visiting Astronomer, Cerro Tololo Inter-American Observatory,
National Optical Astronomy Observatory, which are operated by the Association 
of Universities for Research in Astronomy under cooperative agreement with 
the National Science Foundation.
}

\date{Received; accepted}

\abstract{
We present a model atmosphere analysis of cool hydrogen-rich white dwarfs
observed at the Very Large Telescope (VLT) with the X-shooter spectrograph.
The intermediate-dispersion and high signal-to-noise ratio of the spectra allowed us
to conduct a detailed analysis of hydrogen and heavy element line profiles. In particular,
we tested various prescriptions for hydrogen Balmer line broadening parameters and
determined the effective temperature and surface gravity of each star. 
Three objects (NLTT~1675, 6390 and 11393) show the presence of heavy elements (Mg, Al, Ca, or Fe).
Our abundance analysis revealed a relatively high iron to calcium ratio in NLTT~1675 and NLTT~6390.
We also present an analysis of spectropolarimetric data obtained at the VLT using
the focal reducer and low dispersion spectrograph (FORS) and we established strict upper limits
on the magnetic field strengths in three of the DAZ white dwarfs and determined the longitudinal magnetic
field strength in the DAZ NLTT~10480. The class of DAZ white dwarfs comprises objects that are possibly accreting 
material from their immediate circumstellar environment and the present study contributes in 
establishing class properties.
}

\keywords{white dwarfs -- stars: atmospheres -- stars: abundances -- stars: magnetic field}

\titlerunning{Chemical composition of cool white dwarfs}
\authorrunning{A. Kawka \and S. Vennes}

\maketitle

\section{Introduction}

Studies of heavy elements in cool hydrogen-rich (DA) white dwarfs were originally
restricted to a few cases, i.e., the cool white dwarf G74-7 \citep{lac1983} and 
the ZZ Ceti star G29-38 \citep{koe1997}, because of their rarity. Diffusion timescales in
cool DA white dwarfs are relatively short \citep{paq1986}, and
heavy elements were not expected to be detectable in the photosphere of these objects. 
Therefore, heavy elements such as calcium in G74-7 must be continuously supplied from an external source.
Incidentally, \citet{zuc1987} reported an infrared excess in G29-38, and \citet{gra1990} showed 
that the excess is caused by the presence of circumstellar dust. The detection of heavy 
elements (Mg, Ca, Fe) in the atmosphere of G29-38 by \citet{koe1997} strengthened the link
between atmospheric contaminants and the presence of circumstellar material.

Surveys of white dwarfs conducted using high-dispersion spectrographs attached to large aperture 
telescopes \citep[e.g.,][]{zuc1998,zuc2003,koe2005} reveal
that, contrary to earlier assessments, a significant fraction of DA white dwarfs, labelled DAZ white dwarfs,
are contaminated with elements heavier than helium. Within a sample of $\sim 100$ DA white dwarfs that 
are not in close binary systems, \citet{zuc2003} showed that $\approx 25$\% of the stars 
exhibit photospheric heavy element lines; for the majority of
stars in their sample only calcium lines are observed, but in some cases
other elements such as magnesium, iron, aluminium and silicon are also detected.

Our study of cool white dwarfs in the revised NLTT catalogue \citep{sal2003}
unveiled new cases of external contaminations. 
The stars were selected using a reduced proper motion diagram combined with optical
and infrared colours \citep{kaw2004} resulting in a catalogue of $\approx 400$
objects; only about half of these objects had previously been spectroscopically 
confirmed as white dwarfs. 
Although the NLTT catalogue is deemed incomplete at low Galactic latitude and far south \citep[see][]{lep2005}, 
it still contains many stars that remain largely unstudied. More complete catalogues utilizing the Digital
Sky Surveys, such as the LSPM catalogue of stars with proper motions greater 
than 0.15'' yr$^{-1}$ \citep{lep2005} have been compiled and complement 
the NLTT catalogue where it is incomplete. 

\citet{kaw2006} conducted low-dispersion spectroscopic observations 
of white dwarf candidates from the NLTT catalogue and listed
49 new objects including 30 DA white dwarfs, three DAZ
white dwarfs including NLTT~43806, and 16 non-DA white dwarfs.
In following-up on NLTT~43806,
\citet{zuc2011} obtained high signal-to-noise and high-dispersion spectra showing it
to harbour a weak magnetic field. 

In this work, we present intermediate-dispersion and high signal-to-noise (S/N) 
ratio spectroscopy and a model atmosphere analysis of new DAZ white dwarfs
that were selected from the New Luyten Two-Tenths (NLTT) catalogue. 
These new objects add to the few hydrogen-rich white dwarfs that have been 
scrutinized for heavy elements other than calcium. 
We presented a first report on this programme with the analysis of the coolest DAZ in 
our sample \citep[NLTT~10480;][]{kaw2011a}. The X-shooter spectrum of
NLTT~10480 showed Zeeman-split calcium and hydrogen lines that revealed
a surface magnetic field of $B_s = 0.519\pm0.004$ MG. At the present time, only three other magnetic
DAZ white dwarfs are known: G77-50 \citep{far2011}, NLTT~43806 \citep{zuc2011}
and LTT~8381 \citep{koe2009}. 
Apart from these four magnetic DAZ white dwarfs, two
more white dwarfs with helium dominated atmospheres that are contaminated
with heavy elements have also been shown to be weakly magnetic: G165-7 
\citep{duf2006} and LHS~2534 \citep{rei2001}. Recently, \citet{pot2010} and 
\citet{nor2011} have proposed that magnetic fields in white dwarfs may be generated
during the common-envelope phase of interacting binaries. Applying this model to the
case of G77-50,
\citet{far2011} proposed that the presence of
a weak field in this star may be the result of a past common-envelope episode
with a planetary component during the white dwarf's formative years.

In the following Sect. 2
we describe observations obtained at the European Southern Observatories 
(ESO) using the New Technology Telescope (3.6-m) and the Very Large Telescopes 
(VLTs) as well as the 4-m telescope at Cerro Tololo Inter-American Observatory
(CTIO). Sect. 3 presents our model atmosphere analysis including details of the 
model structure (Sect. 3.1) and heavy-element line profile (Sect. 3.2) calculations. Using these models
we determine the effective temperature and surface gravity for each star (Sect. 3.3),
and measure their heavy element abundance (Sect. 3.4). The stellar radial
velocities and kinematics, and estimates of the magnetic field strengths are presented in Sect. 3.5 and
Sect. 3.6, respectively. We summarize and discuss some implications of our results in Sect. 4.

\section{Observations}

\subsection{Spectroscopy}

\begin{table}
\caption{Log of spectroscopic observations\label{tbl-log}}
\centering
\begin{tabular}{rllc}
\hline\hline
NLTT & Instrument & UT Date & Exp. Time \\
     &            &         &  (s)      \\
\hline\\
1675  & NTT/EFOSC2    & 2009 Aug. 24 & $2\times1500$ \\
      & VLT/X-shooter & 2010 Nov. 13 & $2\times2400$ \\
      &               & 2010 Dec. 8  & $2\times2400$ \\
6390  & VLT/FORS1     & 2007 Nov. 1  & $2\times1500$ \\
      & VLT/X-shooter & 2011 Jan. 1  & $1\times2400$ \\
11393 & VLT/FORS1     & 2007 Nov. 3  & $2\times1260$ \\
      & NTT/EFOSC2    & 2008 Oct. 21 & $2\times1200$ \\
      & VLT/X-shooter & 2010 Nov. 13 & $2\times2400$ \\
      &               & 2011 Jan. 11 & $2\times2400$ \\
23966 & CTIO/R-C Spec & 2008 Feb. 25 & $2\times2400$ \\
      & VLT/FORS2     & 2010 Jan. 23 & $4\times1200$ \\
      & VLT/X-shooter & 2011 Jan. 6  & $1\times1800$ \\
      &               & 2011 Jan. 30 & $1\times1800$ \\
      &               & 2011 Mar. 26 & $1\times1800$ \\
\hline
\end{tabular}
\end{table}

\begin{figure}[t!]
\includegraphics[width=1.00\columnwidth]{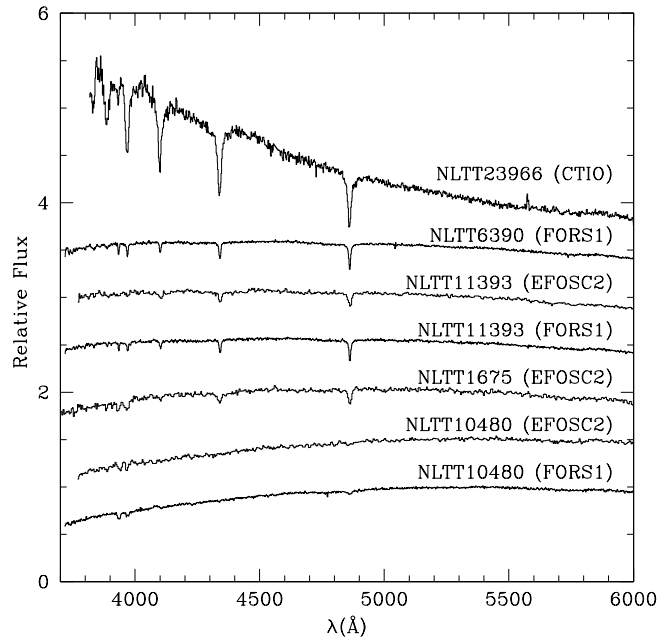}
\caption{Identification spectra, $f_\lambda$ versus $\lambda$, normalized at $\lambda=5500$\AA\ and shifted up for
clarity.
\label{fig-spec}}
\end{figure}

\begin{figure}
\includegraphics[width=1.00\columnwidth]{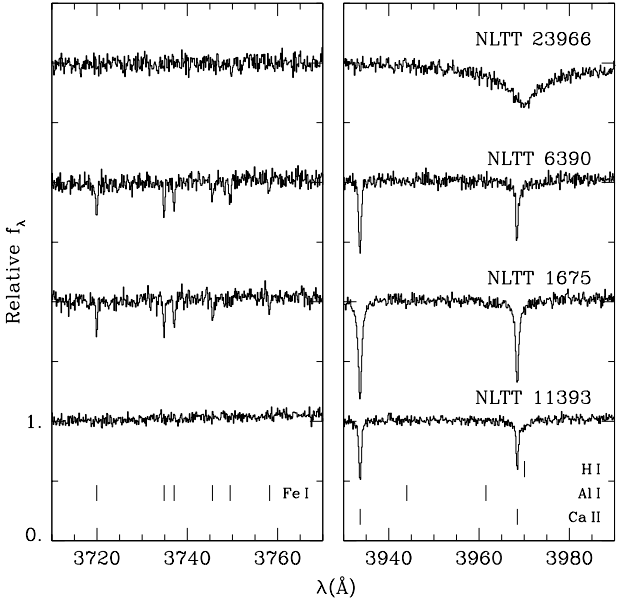}
\caption{X-shooter spectra covering iron ({\it left}) and calcium ({\it right}) 
lines in four cool, hydrogen-rich white dwarfs from the NLTT catalogue. 
Photospheric calcium lines are observed in three stars, while iron is only 
observed in NLTT~1675 and 6390. Contrary to our initial report 
\citep{kaw2011b}, our new X-shooter spectra show that NLTT~23966 is devoid of 
calcium. The possible identification of aluminium in reported for NLTT~1675.
\label{fig-cafe}}
\end{figure}

We first observed objects from the present sample during a programme aimed
at identifying new white dwarfs in the NLTT catalogue. 
NLTT~6390 and NLTT~11393 were 
observed with the focal reducer and low dispersion spectrograph (FORS1) 
attached to the 8m UT2 (Kuenyen) at Paranal Observatory on UT 2007 November 1 
and 3 as part of our spectropolarimetric 
survey of white dwarfs. The purpose of these observations was to confirm 
the nature of the observed sample of stars and to 
search for weak magnetic fields. We used the 600B grism combined with a 
slit-width of 1 arcsecond that provided a resolution of 6 \AA\ and a spectral 
range of 3780 to 6180 \AA. The observations consisted of a sequence of two 
consecutive exposures. In the first exposure the Wollaston prism is rotated to 
$-45^\circ$ followed by a second exposure with the Wollaston prism rotated to 
$+45^\circ$ from which we extracted the flux and circular polarization spectra.

As part of our identification programme
we observed NLTT~23966 with the R-C 
spectrograph attached to the 4m telescope at CTIO
on UT 2008 February 25. We used the KPGL2 (316 lines per mm) 
grating with the WG360 filter to block out the second 
order contamination. We set the slit-width to 1.5 arcseconds which provided a 
spectral resolution of $\sim 8$ \AA\ and a spectral range of 3820 to 7500 \AA.
The spectra were flux calibrated with the flux standard GD~108.
We also selected a sample of spectroscopically confirmed hydrogen-rich white dwarfs 
from the NLTT catalogue, including NLTT~23966, for spectropolarimetric observations with the aim
of searching for weak magnetic fields. 
NLTT~23966 was observed with the FORS2 attached to the 8m UT1 (Antu) at Paranal
Observatory on UT 2010 Jan 23. We used the 1200R grism with the order
blocking filter GG435 and 2kx4k MIT CCD.
The slit-width was set to 1 arcsecond which resulted in a spectral resolution
of $\sim 3$ \AA\ and a spectral range of 5810 to 7290 \AA. The observations
were carried out in the same way as for NLTT~6390 and NLTT~11393 with the
FORS1 spectrograph with one sequence consisting of two consecutive exposures 
with the Wollaston prism rotated 90$^\circ$ between the two exposures. Two 
sequences were obtained for NLTT~23966.

NLTT~1675 was first observed with the ESO Faint Object Spectrograph and Camera
(EFOSC2) attached to the 3.6m New Technology Telescope (NTT) at La Silla
Observatory on UT 2009 August 24. We used Grism 11 which has 300 lines per mm 
and a blaze wavelength of 4000 \AA. The slit-width was set to 1 arcsecond which 
resulted in a spectral resolution of $\sim 14$ \AA\ and a spectral range of 
3700 to 7250 \AA. We have also observed NLTT~11393 using this setup on
UT 2008 October 21. The observations were carried out at the parallactic angle
and were flux calibrated with the flux standard Feige 110.

\begin{table}[t!]
\caption{Line identifications\label{tbl-line}}
\centering
\begin{tabular}{llccc}
\hline\hline
Ion & $\lambda$ & \multicolumn{3}{c}{E.W.} \\
    & (\AA)     & \multicolumn{3}{c}{(m\AA)} \\
\hline
              &         & \multicolumn{3}{c}{NLTT} \\
             &          & 1675 & 6390 & 11393 \\
\cline{3-5}\\
\ion{Fe}{i}  & 3440.606 & 148  & 141  & ... \\ 
\ion{Fe}{i}  & 3565.379 & 38 : & 48   & ... \\ 
\ion{Fe}{i}  & 3570.097 & 80   & 109  & ... \\
\ion{Fe}{i}  & 3581.195 & 185  & 161  & ... \\
\ion{Fe}{i}  & 3608.859 & 84 : & ...  & ... \\
\ion{Fe}{i}  & 3618.768 & 97 : & ...  & ... \\
\ion{Fe}{i}  & 3631.463 & 104 :& 95   & ... \\
\ion{Fe}{i}  & 3647.842 &  ... & 46   & ... \\
\ion{Fe}{i}  & 3719.935 & 117  & 191  & ... \\
\ion{Fe}{i}  & 3734.864 & 150  & 119  & ... \\
\ion{Fe}{i}  & 3737.131 & 128  & 105  & ... \\
\ion{Fe}{i}  & 3745.561 & 112  & 108  & ... \\
\ion{Fe}{i}  & 3748.262 & ...  &  50  & ... \\
\ion{Fe}{i}  & 3749.485 &  71 :& 106  & ... \\
\ion{Fe}{i}  & 3758.233 & 53   & 72   & ... \\
\ion{Fe}{i}  & 3798.511 & 46   & ...  & ... \\
\ion{Fe}{i}  & 3815.840 & 42 : & 29 : & ... \\
\ion{Fe}{i}  & 3820.425 & 115  & 117  & ... \\
\ion{Fe}{i}  & 3825.880 & 46   & 63   & ... \\
\ion{Mg}{i}  & 3832.304 & 96 : & 60 : & ...\\
\ion{Fe}{i}  & 3834.222 & 42   & ...  & ... \\
\ion{Mg}{i}  & 3838.292 & 58 : & 91 : & ...\\
\ion{Fe}{i}  & 3859.911 & 119  & 77   & ... \\
\ion{Ca}{ii} & 3933.663 & 1161 & 485  & 420    \\ 
\ion{Al}{i}  & 3944.006 & 48 : & ...  & ... \\
\ion{Al}{i}  & 3961.520 & 36 : & ...  & ... \\
\ion{Ca}{ii}\tablefootmark{1} & 3968.469 & 866  & 499  & 420  \\  
\ion{Fe}{i}  & 4045.813 & 56   & ...  & ... \\
\ion{Ca}{i}  & 4226.728 & 73   & ...  & ... \\   
\ion{Fe}{i}  & 4383.544 & 51   & 31   & ... \\   
\ion{Ca}{ii} & 8542.09  & 64   & ...  & ... \\  
\ion{Ca}{ii} & 8662.14  & 45   & ...  & ... \\  
\hline
\end{tabular}\\
\tablefoottext{1}{Blended with weak H$\epsilon$.}
\end{table}

Figure~\ref{fig-spec} shows the low-dispersion identification spectra of the 
five cool hydrogen rich white dwarfs for which higher dispersion spectra
were obtained with VLT/X-shooter, including the identification 
spectra of the cool magnetic DAZ NLTT~10480 which was analyzed by 
\citet{kaw2011a}. The low dispersion spectra of these stars showed the 
presence of the \ion{Ca}{ii}$\lambda3933$\AA\ line and therefore were selected for
follow-up high-dispersion observations with X-shooter. In the case of NLTT~23966
the \ion{Ca}{ii}$\lambda3933$\AA\ line seen in the CTIO spectrum was the result of 
background noise.

Following-up on our low-dispersion spectroscopic observations, 
we obtained a set of echelle spectra for all five objects (including
NLTT~10480) using the X-shooter spectrograph attached to the UT2 (Kueyen) at 
Paranal Observatory \citep{ver2011}. The spectra were obtained with the slit-width set to 0.5, 
0.9 and 0.6 arcsecond for the UVB, VIS and NIR arms, respectively. This set-up 
delivered a resolving power of 9100 for UVB, 8800 for VIS and 6200 for NIR. 

Figure~\ref{fig-cafe} shows X-shooter spectra of NLTT~1675, 6390, 11393 and
23966. Photospheric lines of calcium, iron and aluminium are
marked. Table~\ref{tbl-log} summarizes our spectroscopic observations and
Table~\ref{tbl-line} lists the line identifications and equivalent width measurements 
in the X-shooter spectra. The observations of NLTT~10480 were originally presented in \citet{kaw2011a}.

\subsection{Photometry}

\begin{table}
\caption{Photometric and astrometric properties of the sample\label{tbl-star}}
\centering
\begin{tabular}{lccrr}
\hline\hline
NLTT & V\tablefootmark{1} & J \tablefootmark{2} & $\mu_\alpha\cos{\delta}$ \tablefootmark{3} & $\mu_\delta$ $^3$ \\
     & (mag) & (mag)           & (mas yr$^{-1}$) & (mas yr$^{-1}$) \\
\hline\\
1675\tablefootmark{4}   & $17.73\pm0.04$ & $16.56\pm0.13$ & $203\pm6$  &  $-100\pm6$  \\
6390  & $17.33\pm0.04$  & $16.36\pm0.13$ & $440\pm6$  &  $-213\pm6$  \\
11393 & $17.16\pm0.04$  & $16.04\pm0.08$ & $250\pm6$  &  $-239\pm6$  \\
23966 & $17.13\pm0.04$  & $16.54\pm0.11$ & $-160\pm6$ & $-51\pm6$    \\
\hline
\end{tabular}\\
\tablefoottext{1}{This work; $V$-magnitude for NLTT~1675 estimated from SDSS $ugriz$ photometry.}
\tablefoottext{2}{2MASS, \citet{skr2006}.}
\tablefoottext{3}{rNLTT, \citet{sal2003}.}
\tablefoottext{4}{Also known as LSPM~J0031+2218, \citet{lep2005}.}
\end{table}

We searched for photometric measurements using VizieR. The Two Micron All Sky 
Survey (2MASS) listed infrared $JHK$ magnitudes, but only the $J$ magnitude
was of acceptable quality. For NLTT~6390, 11393 and 23966 we used the
acquisition images from the X-shooter observations to estimate a $V$ magnitude.
The X-shooter acquisition images for NLTT~1675 were unusable.
To set the zero point for these magnitudes we used 11 acquisition images of 
Feige 110 obtained between UT 2010 Dec 10 and 2011 Jan 1. We employed the 
atmospheric extinction table of \citet{pat2011}. 

For NLTT~1675, we also attempted to
measure a $V$ magnitude from the EFOSC2 acquisition images, however both 
acquisition images appear to have been taken through thin clouds and the final 
magnitudes are unreliable. Fortunately, Sloan Digital Sky Survey (SDSS) photometry
is available for NLTT~1675 and we have used these measurements to determine a $V$ 
magnitude: We obtained a best fit model spectrum to SDSS photometry 
and then we convolved this model with the $V$ bandpass \citep{bes1990} providing
us with an estimate of the $V$ magnitude.

Table~\ref{tbl-star} lists the photometric and proper motion measurements
for each star.

\section{Analysis}

\subsection{Model atmospheres and spectral syntheses}

We extended our grid of model atmospheres for cool hydrogen-rich white dwarfs
employed in \citet{kaw2011a} to encompass the
effective temperature range $4\,900\le T_{\rm eff}\le 8\,000$\,K in 100\,K steps
and the surface gravity range $7.0\le \log{g}\le8.75$ in steps of 0.25 dex.
The models are in convective and radiative equilibrium. 
Again, all relevant species (H, H$^+$,  H$_2$,  H$_2^+$,  H$_3^+$) are included in the statistical 
equilibrium equation, with electrons contributed by identifiable trace elements (e.g., 
calcium) included in the charge conservation equation. 
However, in warmer models that are relevant to the present study ($T_{\rm eff}\ge 6\,000$) electrons
are contributed mostly by the ionization of hydrogen atoms.
The H$_2$-H and H-H collision-induced absorptions in the far Ly$\alpha$ wing
\citep[see][]{kow2006} are included using opacity tables from \citet{roh2011}.
Synthetic colours as well as detailed hydrogen and heavy element line profiles 
are computed using the model structures. Table~\ref{tbl-model} lists some 
photometric properties of these models. The colour indices at shorter 
wavelengths are effected by the Ly$\alpha$ collision-induced absorptions.

The calculation of synthetic hydrogen line profiles deserves further attention.
The FWHM for resonant levels computed by \citet{ali1965,ali1966}, the AG formalism, assuming ``resonance'' 
interaction ($\propto R^{-3}$) is
\begin{equation}
\Gamma_{3,u} \equiv \frac{w}{N} = 4.8624\times10^8\,\Big{(}\frac{g_l}{g_u}\Big{)}^{1/2} \frac{f_{lu}}{\nu_{lu}},
\end{equation}
where, for example, the lower ($l$) and upper ($u$) levels were taken as 1s and 2p in \citet{ali1965,ali1966},
with $g_l/g_u = 1/3$ and the oscillator strength $f_{lu} = 0.4162$, providing a FWHM for the 2p level
of $2.4\times10^{-8}$\,rad\,s$^{-1}$\,cm$^3$.

\begin{figure}[t!]
\includegraphics[width=1.00\columnwidth]{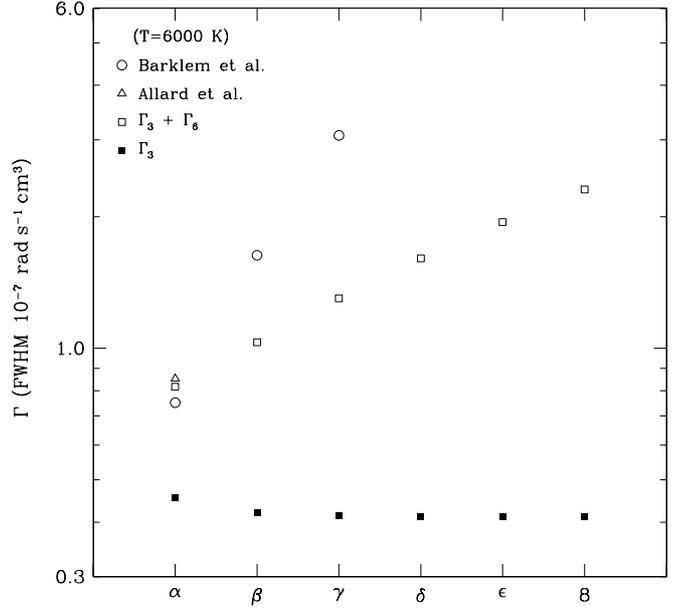}
\caption{Self-broadening parameter $\Gamma$ for members of the Balmer line 
series. The widths ($\Gamma_3$) from \citet{ali1965,ali1966} follow a markedly different 
trend than widths from \citet{bar2000a}. Improvements are obtained by 
adding a van der Waals term ($\Gamma_6$) to \citet{ali1965,ali1966}: $\Gamma_3+\Gamma_6$.
\label{fig-wid}}
\end{figure}

\citet{ber1991} proposed to apply this formalism to principal quantum numbers $l=1$ and $u$ by
taking $g_l/g_u = (l/u)^2$ and $f_{lu}$ as the total oscillator strength between
principal quantum numbers $l=1$ and $u$. This formalism offers a means to 
estimate the total width of the level $u$ that includes the s, p, d 
contributions, i.e., resonant and non-resonant levels. Figure~\ref{fig-wid} 
shows Balmer line widths calculated by summing the lower ($u$) and upper ($u'$) 
level widths $\Gamma_3 = \Gamma_{3,u}+\Gamma_{3,u'}$. In this formalism, the total 
width is dominated by the $u=2$ width, while the contribution of the upper 
level $u'$ to the width decreases with increasing values of $u'$.

A comparison with the calculations of \citet{bar2000a} for H$\alpha$, 
$\beta$, and $\gamma$, and \citet{all2008} for H$\alpha$ shows that the total 
line width is not well represented by this generalisation of the AG formalism.
Figure~\ref{fig-wid} shows that the two formalisms behave differently 
along the line series. Although the \citet{ali1965,ali1966} values are nearly 
constant (when expressed in rad\,s$^{-1}$), the width computed by 
\citet{bar2000a} triples between H$\alpha$ and $\gamma$. The 
\citet{ali1965,ali1966} calculations are dominated by 2p level width and are 
based on the resonance 1s-2p transition, while the widths tabulated by 
\citet{bar2000a} rely on the $n$p-$m$d formalism for the $n-m$ transition 
and their calculations employ the ``dispersive-inductive'' (van der Waals) 
interaction ($\propto R^{-6}$) in addition to the ``resonance'' interaction.

Consequently, the AG formalism may be improved by adopting a van de Waals 
term of the form \citep[see][]{kur1981}
\begin{equation}
\Gamma_6 \approx 8\times10^{-9}\, T_4^{3/10}\, \Big{(}u^4-l^{\,4}\Big{)}^{2/5},
\end{equation}
where $T_4 = T/10^4$. The total line width is then obtained by adding the 
resonance \citep{ali1965,ali1966} and van der Waals terms: 
$\Gamma = \Gamma_3+\Gamma_6$. Figure~\ref{fig-wid} shows that the resulting 
line width increases along the series although the widths calculated by 
\citet{bar2000a} remain considerably larger.

\citet{bar2000a} also determined validity limits 
for the quasi-static framework: 35, 13, and 8 \AA\ for H$\alpha$, $\beta$ and 
$\gamma$, respectively, at $T=4665$\,K, revealing potential difficulties with
higher members of the Balmer line series. 

In order to assess possible systematic errors in the measurement of 
stellar parameters based on Balmer line profiles, we explored three different 
approaches for the calculation of Balmer line spectra. In the first approach, 
the Balmer line profiles from H$\alpha$ to the series limit are computed using 
widths from \citet{ali1965,ali1966} alone, $\Gamma_{\rm AG}$, while in the 
second approach we combined AG and van der Waals widths, 
$\Gamma_{\rm AG,vdW}$. In the last approach, the line profiles from H$\alpha$ 
to H$\gamma$ are computed using widths from \citet{bar2000a}, 
$\Gamma_{\rm BPO}$, multiplied by factors of 0.75 and 1.0, again to 
assess uncertainties, while the upper Balmer lines are computed using
$\Gamma_{\rm AG,vdW}$. The Voigt profiles including Doppler and 
self-broadening are convolved with Stark-broadened profiles from 
\citet{lem1997}.

\subsection{Heavy element line profiles}

The dominant broadening mechanism at $\sim6000$\,K is collision with hydrogen atoms, 
but the effect of Stark broadening is also included as it contributes significantly at $\sim8000$\,K.

For broadening of \ion{Mg}{i}, \ion{Al}{i}, \ion{Ca}{i}, and \ion{Ca}{ii} lines 
due to collisions with hydrogen atoms we employed
the coefficients of \citet{bar2000b} where the full-width at half-maximum 
(FWHM) of the Lorentzian profiles is given by:
\begin{equation}
\frac{w}{n(\ion{H}{i})} = \Big{(}\frac{T}{10^4{\rm K}}\Big{)}^{(1-\alpha)/2}\, \Gamma\ \ {\rm rad\,s^{-1}\,cm^3}.
\end{equation}
For example, we adopted $\log{\Gamma} = -7.562$ for \ion{Ca}{i}$\lambda4226$ and $-7.76$ for 
Ca\,H\&K at $T=10\,000$\,K, and $\alpha = 0.238$ for \ion{Ca}{i}$\lambda4226$ 
and $0.223$ for Ca\,H\&K. 
Stark broadening widths for \ion{Ca}{i} and \ion{Ca}{ii} lines were obtained from
\citet{dim1999} and \citet{dim1992}, respectively. 
Additional parameters (Stark and collisions with hydrogen) were obtained from the compilation of \citet{kur1995}.

Hydrogen molecules provide $\la$20\% of the gas 
pressure in some layers of models at $T_{\rm eff}=$6000~K and a negligible 
fraction at 8000~K. Consequently, the effect of collision with molecules is not 
considered further. 

\subsection{Effective temperature and surface gravity measurements}

Fitting the $V-J$ colour index to synthetic colours (Table~\ref{tbl-model}) 
constrains the effective temperature of the cool white dwarfs 
(Table~\ref{tbl-param}). Within error bars the cooler objects are at
$T_{\rm eff}\approx 6000$\,K, while NLTT~23966 is hotter with 
$T_{\rm eff}\approx 7600$\,K. These preliminary colour-based estimates are 
corroborated by our detailed Balmer line analysis.

\begin{figure}[t!]
\includegraphics[width=1.00\columnwidth]{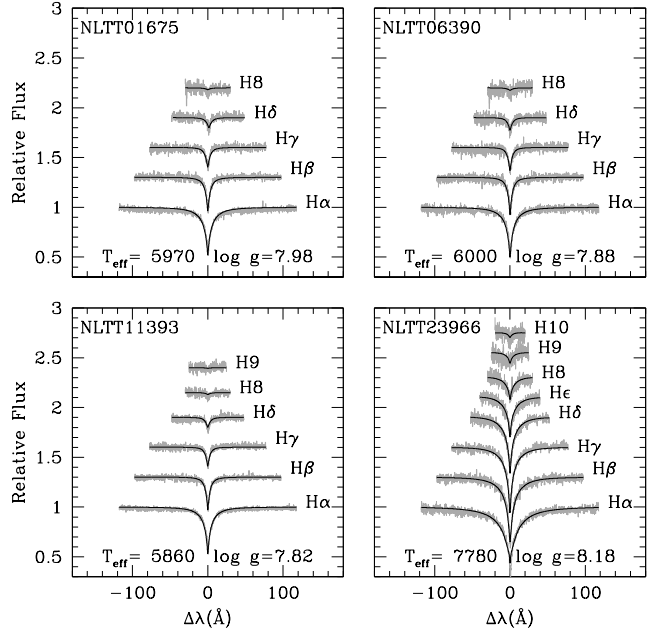}
\caption{Balmer line fits of X-shooter spectra using the line-broadening 
prescription $\Gamma_{\rm AG,vdW}$ (see text). With the exception of NLTT~23966, 
the H$\epsilon$/Ca~H blend is excluded from the line fitting.
\label{fig-balm}}
\end{figure}

Next, we simultaneously constrained $T_{\rm eff}$ and $\log{g}$ by fitting the
X-shooter spectra with the model grids using $\chi^2$ minimization techniques.  
All lines up to the series limit, H8/9 for the cooler stars and H10 for NLTT~23966,
are included in the analysis, but we excluded the H$\epsilon$/Ca~H blend in the three cooler stars.
Figure~\ref{fig-balm} shows the Balmer line profiles of the four white
dwarfs compared to the best-fitting model spectra that were calculated using
$\Gamma_{\rm AG,vdW}$. 

\begin{figure}[t!]
\includegraphics[width=1.00\columnwidth]{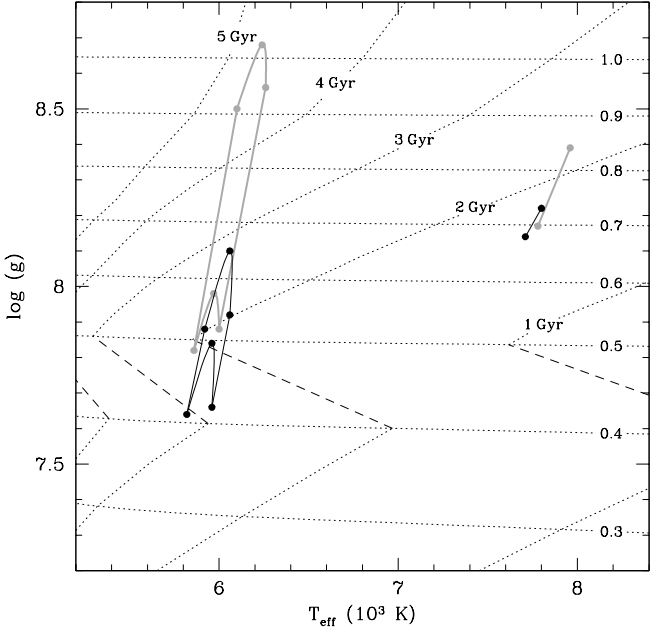}
\caption{Range of parameters ($T_{\rm eff}$ vs. $\log{g}$) determined for a 
sample of four cool hydrogen-rich white dwarfs using three different 
prescriptions for hydrogen line broadening (see text). The measured parameters are compared to 
evolutionary tracks from \citet{ben1999} with masses ranging from 0.3 to 1.0 
$M_\odot$, and ages ranging from 1 to 5 Gyr. An age discontinuity is apparent 
$\sim 0.45\, M_\odot$ because of the different core composition adopted for 
models with $M\le0.4\,M_\odot$ (He) and $M\ge0.5\,M_\odot$ (C/O).
\label{fig-tegr}}
\end{figure}

Figure~\ref{fig-tegr} illustrates the range of effective temperatures and 
surface gravities attained for the sample 
when employing different broadening parameters. Values connected with
thin grey lines were obtained using either $\Gamma_{\rm AG}$ alone (upper points),
or $\Gamma_{\rm AG,vdW}$ (lower points). The values connected with 
full black lines were obtained using $\Gamma_{\rm BPO}$ (H$\alpha,\beta,\gamma$) combined 
with $\Gamma_{\rm AG,vdW}$ for the upper Balmer lines (lower points),
or reduced $0.75\times\Gamma_{\rm BPO}$ values combined with $\Gamma_{\rm AG,vdW}$ (upper points).
The derived parameters are compared to evolutionary cooling tracks \citep{ben1999}
allowing estimates of the mass and cooling age of each object.

The average mass for the three cooler stars computed using 
$\Gamma_{\rm AG,vdW}$ line widths is 0.53\,$M_\odot$. 
Similarly, the average mass is $0.57\,M_\odot$ using the combination
$\Gamma_{\rm BPO}\times0.75$/$\Gamma_{\rm AG,vdW}$.
On the other hand, the average mass of $0.96\,M_\odot$ obtained using $\Gamma_{\rm AG}$ appears improbably high,
while the average mass of $0.44\,M_\odot$ obtained using the combination $\Gamma_{\rm BPO}$/$\Gamma_{\rm AG,vdW}$
appears improbably low.
Higher (lower) values of the surface 
gravity (hence density) are derived as a compensation for using smaller (larger) values 
of $\Gamma$.
Considering the uncertainties in $\Gamma$, it seems appropriate to adopt an
average mass near $0.6\,M_\odot$ \citep[see][]{tre2010} for the three cooler stars.

On the other hand, the 
surface gravity, hence mass, of NLTT~23966 is only marginally affected by 
variations in the self-broadening $\Gamma$ value because of the dominant effect of Stark broadening at 
higher temperatures. The higher than average mass estimated for NLTT~23966 may 
also be attributed to the approximate treatment of convection in current model 
calculations \citep{tre2011}.

Table~\ref{tbl-param} lists the adopted effective temperature and surface 
gravity of the cool white dwarfs. As a compromise, we adopted the average of the values
obtained using model spectra calculated with $\Gamma_{\rm AG,vdW}$ or
$\Gamma_{\rm BPO}\times 0.75$. Individual white dwarf masses were calculated
following \citet{ben1999}.

\begin{table}[t!]
\caption{Stellar properties \label{tbl-param}}
\centering
\begin{tabular}{rcccc}
\hline\hline
NLTT & $T_{\rm eff}$ ($V-J$) & $T_{\rm eff}$ \tablefootmark{1} & $\log{g}$ $^1$ & Mass \\
     &        (K)            &    (K)                & (cm\,s$^{-2}$) & ($M_\odot$)  \\
\hline
1675  &  $5800\pm300$        &  $6020\pm50$           & $8.04\pm0.07$ & $0.61\pm0.04$ \\
6390  &  $6200\pm300$        &  $6040\pm40$           & $7.90\pm0.07$ & $0.53\pm0.04$ \\
11393 &  $5800\pm300$        &  $5890\pm30$           & $7.86\pm0.06$ & $0.51\pm0.03$ \\
23966 &  $7600\pm300$        &  $7790\pm30$           & $8.20\pm0.04$ & $0.72\pm0.03$ \\
\hline
\end{tabular}\\
\tablefoottext{1}{Average of the parameters determined using Balmer line 
profiles computed using $\Gamma_{\rm AG,vdW}$ and $\Gamma_{\rm BPO}\times 0.75$ 
models.}
\end{table}

\subsection{Abundance of heavy elements}

\begin{figure}[t!]
\includegraphics[width=1.00\columnwidth]{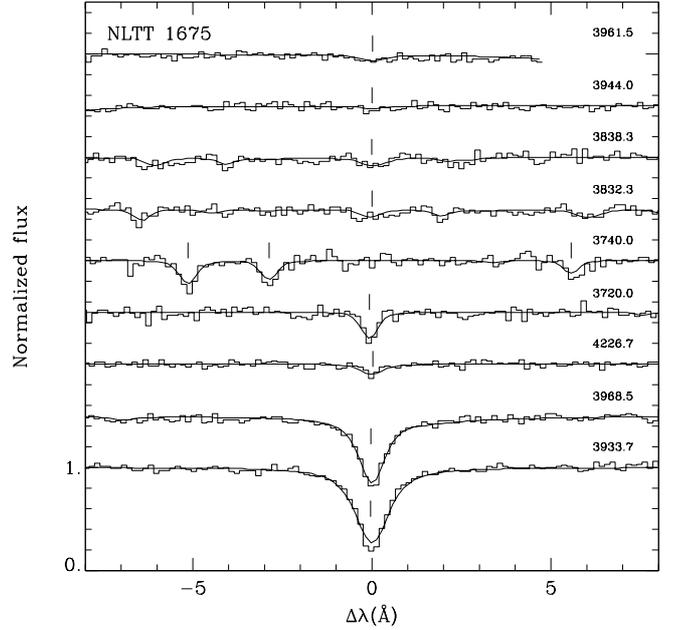}
\caption{X-shooter spectra of NLTT~1675 compared to the best fit model. The
wavelength scale is centred on the values indicated on the right. The top
two spectra show the weak \ion{Al}{i} lines, followed by spectra showing
\ion{Mg}{i} lines around 3838 \AA\ and 3832 \AA\ and \ion{Fe}{i} lines around 
3740 \AA\ and 3720 \AA. Finally, the bottom three spectra show the \ion{Ca}{i} 
and \ion{Ca}{ii} lines. \label{fig-nltt1675}}
\end{figure}

The abundances were computed at $\log{g}=8.0$ and $T_{\rm eff} = 6000\pm100$\,K for the cooler stars,
and at $\log{g}=8.25$ and $T_{\rm eff} = 7800\pm100$\,K for NLTT~23966.
Table~\ref{tbl-abun} lists the measurements obtained by fitting the spectra with models of varying composition; 
the quoted errors include the (mild) effect of varying the temperature by $\pm$100~K, and the upper limits
are taken at 99\% certainty (2.6$\sigma$).
However, the magnesium and aluminium abundances are based on $\sim2\sigma$ detections and higher signal-to-noise
ratio spectra are required to help improve the measurements.
On the other hand, a strict upper limit on the calcium abundance in NLTT~23966 shows that its atmosphere is devoid of
heavy elements. 
Figure~\ref{fig-nltt1675} shows the various metal lines detected in NLTT~1675
compared to the best fitting model spectra.

\begin{table}[t!]
\caption{Abundance measurements \tablefootmark{1} \label{tbl-abun}}
\centering
\begin{tabular}{lcccc}
\hline\hline
NLTT & $\log{\rm Mg/H}$ & $\log{\rm Al/H}$ & $\log{\rm Ca/H}$ & $\log{\rm Fe/H}$ \\
\hline
1675  & $-$8.56$\pm$0.12 & $-$9.28$\pm$0.17 & $-$9.53$\pm$0.03  &   $-$8.63$\pm$0.13 \\
6390  & $-$8.66$\pm$0.20 & $<-$9.2          & $-$10.00$\pm$0.04 &   $-$8.57$\pm$0.11 \\
11393 & $<-8.7$         &  $<-$9.4         &  $-$10.24$\pm$0.04 &   $<-$9.4 \\
23966 & ...             & ...              &  $<-10.4$      &   ...              \\
\hline
\end{tabular}
\tablefoottext{1}{Error bars are 1$\sigma$ and the upper limits are given at 99\% confidence.}
\end{table}

\subsection{Radial velocity measurements, kinematics, and ages}

We measured radial velocities $\varv_r=15.0, 83.2, 64.2$ and 
$66.4$\,km\,s$^{-1}$ for NLTT~1675, 6390, 11393 and 23966, respectively. We 
estimated the error to be $5.0$\,km\,s$^{-1}$ based on the scatter in 
individual line measurements and the expected precision of a $R\sim9000$ 
spectrograph. Taking into account the expected gravitational redshift of the 
white dwarfs (based on the spectroscopically determined parameters), the actual 
velocities of the stars are $\varv = -16.5\pm5.8, 58.3\pm5.8, 40.9\pm5.4$ and 
$25.4\pm5.4$\,km\,s$^{-1}$ for NLTT~1675, 6390, 11393 and 23966, respectively.

Table~\ref{tbl-kin} lists the cooling age, absolute magnitude, distance
and Galactic space velocities of the sample. The ages were computed using
the evolutionary models of \citet{ben1999} and the absolute magnitudes were
calculated using our model grid and radii from \citet{ben1999}. The 
photometric distances were calculated using the apparent
(Table~\ref{tbl-star}) and absolute (Table~\ref{tbl-kin}) Johnson V magnitudes. Finally, the Galactic velocity
components $UVW$ were calculated using \citet{joh1987}. The lag in $V$ 
suggest that the white dwarfs belong to the old thin disk \citep{sio1988,pau2003,pau2006}.

\begin{table}[t!]
\caption{Age, distance and kinematics \label{tbl-kin}}
\centering
\begin{tabular}{lcccc}
\hline\hline
NLTT  & Age     &      $M_V$      &   $d$    &   $U,V,W$      \\
      &  (Gyr)  &      (mag)      &  (pc)    & (km\,s$^{-1}$) \\
\hline
1675  & 2.2-2.9 & 14.27$\pm$0.11  & 49$\pm$3 & $-$15$\pm$6,$-$44$\pm$8, $-$4$\pm$6  \\
6390  & 1.7-2.3 & 14.06$\pm$0.11  & 45$\pm$2 & $-$37$\pm$5,$-$101$\pm$9,$-$25$\pm$8  \\
11393 & 1.8-2.2 & 14.13$\pm$0.10  & 40$\pm$2 &     2$\pm$5,$-$72$\pm$7, $-$5$\pm$7  \\
23966 & 1.6-1.9 & 13.45$\pm$0.06  & 54$\pm$2 & $-$23$\pm$5,$-$31$\pm$6, $-$6$\pm$6  \\
\hline
\end{tabular}
\end{table}

\subsection{Magnetic field strengths}

A lack of splitting at the instrument resolution ($R\sim 9000$) imposes
a limit on the magnetic field strength. Setting a limit of approximately
1/3 of a resolution element to the putative magnetic broadening of narrow line cores
due to Zeeman splitting:
\begin{equation} 
\Delta\lambda \equiv k\,B_s \approx \frac{1}{3}\frac{\lambda}{R}
\end{equation}    
where $k=4.67\times10^{-13}\lambda^2\,g_{\rm eff}$, $\lambda$ is the wavelength in \AA, $B_s$ is the average surface field in G, and $g_{\rm eff}$ is the effective Land\'e factor ($=\frac{7}{6}$ and $\frac{4}{3}$ for Ca~K and H, respectively).
For Ca~K the resulting limit on the field strength is $B\la$\,17~kG.
However, this effect could be confused with the effect of stellar rotation.

We obtained VLT/FORS spectropolarimetry for three stars (NLTT~6390, 11393 and 23966) 
in addition to NLTT~10480. We calculated the mean longitudinal 
magnetic ($B_l$ in G) field using the weak-field approximation
\citep{ang1973}
\begin{equation}
k\,B_l = \frac{\varv F}{dF/d\lambda}
\label{eqn_bl}
\end{equation}
where $\varv = V/I$ is the degree of circular polarization, 
$F\equiv I$ is the total spectral flux, and $dF/d\lambda$ is the 
flux gradient. First, we fitted the observed Balmer line profiles with model spectra
to determine the best fit line profiles. We then used these line profiles
to calculate the flux gradient and determine the longitudinal magnetic field.
To measure the longitudinal field strength for NLTT~6390 and NLTT~11393 we 
used H$\beta$ to H$\delta$ and for NLTT~23966 we used H$\alpha$.

Table~\ref{tbl-mag} lists the longitudinal field measurements for
three stars (NLTT~6390, 11393 and 23966) from this paper and for the cool
white dwarf NLTT~10480 \citep{kaw2011a}. 
The quoted errors are $1\sigma$. Using the Balmer lines, no fields stronger than $\sim 5$ to $\sim 20$ kG are
detected in NLTT~6390, 11393 and 23966, but we
measured a longitudinal field strength 
$B_l = -200.4\pm124.0$ kG in NLTT~10480, which
is barely a $2\sigma$ detection. The large uncertainty in $B_l$
is mostly due to the weakness of the Balmer lines (H$\beta$, $\gamma$ and $\delta$) in NLTT~10480, but
a measurement based on the calcium lines appears more promising.

\begin{figure}[t!]
\includegraphics[width=1.00\columnwidth]{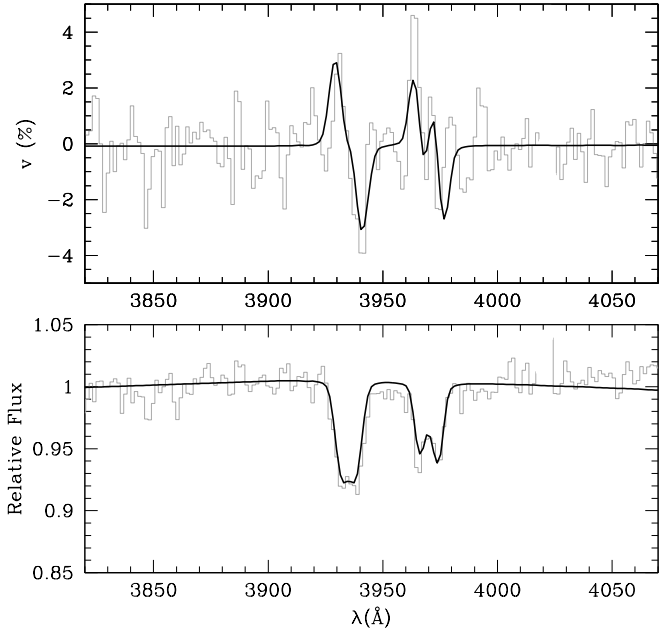}
\caption{Circular polarization ({\it top}) and flux ({\it bottom}) spectra
of NLTT~10480. The observed circular polarization spectrum is compared to a 
model polarization spectrum at $B_l =212$ kG ({\it top}).
\label{fig-nltt10480_ca}}
\end{figure}

Equation~\ref{eqn_bl} includes only the first term of a Taylor
expansion that is used to calculate $\varv$. This is valid where the Zeeman
splitting is small compared to the line width. For Ca~H\&K
in NLTT~10480, the Zeeman splitting is fully resolved and therefore we need to 
include the second term of the Taylor expansion as described in \citet{mat1986}
\begin{equation}
\varv = \frac{1}{F}\,\Big(k\,B_l\,\frac{dF}{d\lambda} + \frac{1}{6}\,(k\,B_l)^3\, \frac{d^3 F}{d\lambda^3}\Big).
\label{eqn_bl2}
\end{equation}
By fitting the Ca\,H\&K circular polarization spectrum depicted in 
\citet{kaw2011a} we measured a longitudinal field of $-212\pm50$ kG in 
NLTT~10480. Figure~\ref{fig-nltt10480_ca} shows the best fit model circular 
polarization and flux spectra to the observed FORS spectra of NLTT~10480.
A reasonable match is achieved between the model and observed spectra, except
for the left $\sigma$ component of the CaII $\lambda 3968$ \AA\ line, which 
appears stronger than the model.
This is an improvement over the Balmer line measurements, but it still has a 
significantly larger uncertainty than the measurements obtained for the other 
white dwarfs in the sample. This is mainly due to the
lower signal-to-noise ratio in the spectral region containing the \ion{Ca}{ii} lines.
Combining the measurement of the surface field of $519\pm4$ kG \citep{kaw2011a}
and the Ca~H\&K measurement of the longitudinal field in NLTT~10480
suggests an inclination below $59^\circ$, in agreement with an angle of
$60\pm3^\circ$ obtained from the strength of the calcium Zeeman-split lines
determined in \citet{kaw2011a}.

\begin{table}[t!]
\caption{Magnetic field measurements\label{tbl-mag}}
\centering
\begin{tabular}{lcl}
\hline\hline
NLTT & $B_l$      & Lines \\
     &  (kG)      &        \\
\hline
6390  & $-4.8\pm9.0$     & H$\beta$, H$\gamma$, H$\delta$ \\
10480 & $-200\pm124$     & H$\beta$, H$\gamma$            \\
      & $-212\pm50$      & Ca\,H\&K                         \\
11393 & $+7.8\pm12.2$    & H$\beta$, H$\gamma$, H$\delta$ \\
23966 & $-2.6\pm3.1$     & H$\alpha$                      \\
\hline
\end{tabular}
\end{table}

\section{Summary and discussion}

We have presented a detailed model atmosphere analysis of high-quality spectroscopy
of a sample of cool DA white dwarfs. We show that the atmospheres of NLTT~1675,
NLTT~6390, and NLTT~11393 are contaminated with heavy elements. On the other
hand we show that, contrary to the DAZ white dwarf NLTT~10480, none of these three new DAZ white dwarfs
harbour a magnetic field. We also refined our analysis of the magnetic
field in NLTT~10480 and confirm our original results \citep{kaw2011a}.

\begin{figure}[t!]
\includegraphics[width=1.00\columnwidth]{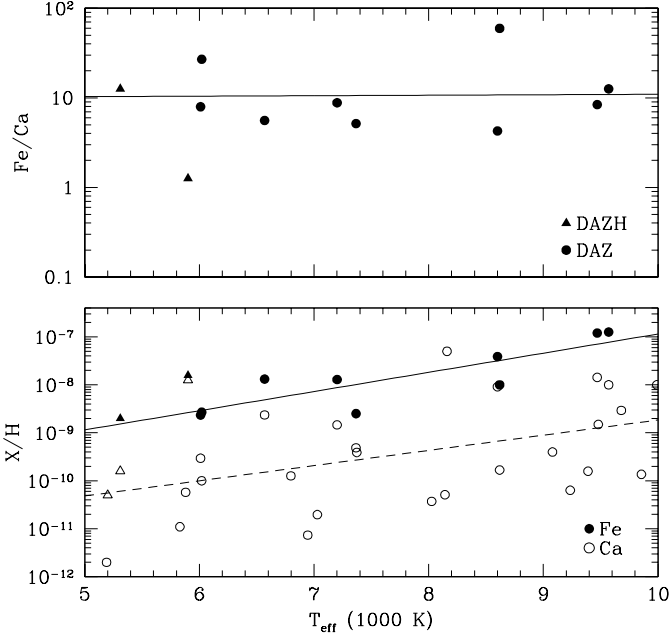}
\caption{Abundance ratio of iron with respect to calcium versus the effective
temperature ({\it top}). The best fitting line that excludes NLTT~43806 is 
also shown. The abundance of iron and calcium versus the effective temperature
({\it bottom}), upper limits are not shown. The best fitting lines for iron 
({\it full line}) and calcium ({\it dashed line}) are shown. \label{fig_abun}}
\end{figure}

Figure~\ref{fig_abun} shows the abundances of iron and calcium relative to hydrogen for
a sample of cool DAZ white dwarfs, including those from the present sample, and the corresponding abundance ratio of
iron with respect to calcium (Fe$/$Ca). The published abundance measurements were taken from 
\citet{zuc2003} and \citet{kaw2011b}, and from \citet{far2011}, \citet{kaw2011a} and \citet{zuc2011} for the three
magnetic white dwarfs, G77-50, NLTT~10480 and NLTT~43806, respectively.
Table~\ref{tbl-cool} lists the selected measurements.
The Fe$/$Ca abundance ratio appears to be constant as a function of temperature
with an average of $<$Fe$/$Ca$> = 13.1$ and a dispersion of $\sigma = 0.33$ dex. 
The linear fit to the data excludes NLTT~43806, which has a Fe$/$Ca abundance 
ratio of $\approx 1.3$ \citep{zuc2011} making it significantly iron poor 
compared to the rest of the sample. Similarly, Fe$/$Al is $\approx 0.6$ in 
NLTT~43806 while it is $\approx 4.5$ in NLTT~1675. Both calcium 
and iron abundances show an increase as a function of temperature, but the scatter of 
iron abundance measurements ($\sigma \sim 0.31$ dex) is lower than that of 
calcium ($\sigma \sim 0.95$ dex). It is possible that higher signal-to-noise 
and higher dispersion spectra of objects apparently devoid of iron would
uncover even lower iron abundances and thereby increase the scatter in the measurements. In summary,
the Fe$/$Ca abundance fraction is $\sim$10 compared to $\approx14$ in the Sun. The Mg$/$Ca abundance fraction of
NLTT~1675, NLTT~6390, G77-50 and NLTT~43806 range between $\approx 6$ and 
$\approx 30$.

\begin{table}
\caption{Inferred accretion rates $\log{\dot{M_X}}$(g s$^{-1}$)
\label{tbl-acc}}
\centering
\begin{tabular}{lcccc}
\hline\hline
NLTT  & Mg      &    Al   & Ca   & Fe \\
\hline
 1675 & 6.59    & 5.97    & 5.98 & 7.17 \\
 6390 & 6.46    & $<$6.02 & 5.48 & 7.20 \\
11393 & $<$6.36 & $<$5.76 & 5.18 & $<$6.31 \\
23966 & ...     & ...     & $<$5.08 & ... \\
\hline
\end{tabular}
\end{table}

We calculated the presumed accretion rates assuming steady state between accretion and
diffusion for magnesium, aluminium, calcium and iron. We used the 
diffusion timescales and mass of the convective zone of \citet{koe2009b}. 
For aluminium we used the diffusion timescale $\log{\tau_{\rm Al}} = 4.23$ 
that \citet{zuc2011} adopted for NLTT~43806, which is a star with
temperature and surface gravity similar to our sample of DAZ white dwarfs. Table~\ref{tbl-acc}
lists the accretion rates for the NLTT~1675, 6390 and 11393. For objects 
with upper limits on abundances, such as NLTT~23966, we list the corresponding upper limits
on accretion rates. The observed abundance ratios relative to calcium
in meteorites (C1 chondrites) or in the Sun \citep{gre2007} are $\log{\rm Mg/Ca}, \log{\rm Al/Ca}, \log{\rm Fe/Ca}=$
1.23, 0.10, and 1.15, respectively. Our sample appears to conform with these ratios. 
Following sample trends, magnesium and iron may well be present in NLTT~11393 but slightly below 
detection limits in our spectra.

Considering the whole sample, the observed calcium and iron abundances follow the excepted 
temperature trend assuming a constant accretion rate over cooling ages \citep[see][]{koe2006},
although it appears somewhat steeper. The calcium and iron abundances
increase by 0.32 and 0.4 dex per 1000~K interval, respectively, compared to
predicted increases of 0.26 and 0.27 dex per 1000~K interval assuming uniform accretion over the sample.
This steeper trend suggests that the accretion flow onto white dwarfs may be slightly abating with time, or that
the actual particle flux at the bottom of the convection zone may be underestimated in cooler, i.e., 
older white dwarfs.

Prior to this study, five DAZ white dwarfs with $T_{\rm eff} \la 6000$\,K) were
known where two of these, NLTT~43806 \citep{kaw2006} and NLTT~10480 
\citep{kaw2011a} were discovered as part of our spectroscopic observations of
white dwarf candidates from the rNLTT catalogue. In this paper we add three more
cool DAZ white dwarfs to this important sample of accreting objects. Also notable is the high incidence of magnetism among this
small sample of stars: Three out of the eight DAZ white dwarfs cooler than $\approx6000$\,K (G77-50, 
NLTT~43806 and NLTT~10480) are weakly magnetic. This incidence is markedly higher than previously
estimated \citep{kaw2007,lie2003}.

\begin{acknowledgements}

S.V. and A.K. are supported by GA AV grant numbers IAA300030908 and IAA301630901, respectively, and by GA \v{C}R grant number P209/10/0967.
A.K. also acknowledges support from the Centre for Theoretical
Astrophysics (LC06014). We thank the referee, D. Koester, for helpful suggestions.
This research has made use of the VizieR catalogue access tool, CDS, Strasbourg, France.
This publication makes use of data products from the Two Micron All Sky Survey,
which is a joint project of the University of Massachusetts and the Infrared
Processing and Analysis Center/California Institute of Technology, funded by
the National Aeronautics and Space Administration and the National Science
Foundation.

\end{acknowledgements}

\appendix

\clearpage

\section{Synthetic Johnson and Sloan colours}

\begin{table}
\caption{Selected synthetic colours \label{tbl-model}}
\centering
\begin{tabular}{lcccrrr}
\hline\hline
$\log{g}$ & $T_{\rm eff} (K)$ & $B-V$ & $V-J$ & $g-J$ & $g-r$ & $r-i$ \\
(cm\,s$^{-2}$) &  (K)          & (mag) & (mag) & (mag) & (mag) & (mag) \\
\hline
7.5 & 5\,800 & 0.535 & 1.149 & 1.428 & 0.396 &  0.141 \\
    & 6\,000 & 0.496 & 1.070 & 1.325 & 0.354 &  0.119 \\
    & 6\,200 & 0.463 & 0.998 & 1.232 & 0.316 &  0.100 \\
    & 6\,400 & 0.432 & 0.931 & 1.145 & 0.281 &  0.081 \\
    & 6\,600 & 0.405 & 0.867 & 1.063 & 0.248 &  0.063 \\
    & 6\,800 & 0.380 & 0.807 & 0.986 & 0.218 &  0.047 \\
    & 7\,000 & 0.358 & 0.750 & 0.913 & 0.190 &  0.031 \\
    & 7\,200 & 0.337 & 0.695 & 0.845 & 0.164 &  0.017 \\
    & 7\,400 & 0.319 & 0.642 & 0.779 & 0.140 &  0.003 \\
    & 7\,600 & 0.303 & 0.591 & 0.716 & 0.117 & $-$0.011 \\
    & 7\,800 & 0.289 & 0.542 & 0.656 & 0.095 & $-$0.024 \\
    & 8\,000 & 0.276 & 0.494 & 0.598 & 0.075 & $-$0.037 \\
8.0 & 5\,800 & 0.537 & 1.139 & 1.419 & 0.395 &  0.139 \\
    & 6\,000 & 0.496 & 1.061 & 1.316 & 0.352 &  0.118 \\
    & 6\,200 & 0.463 & 0.992 & 1.226 & 0.314 &  0.100 \\
    & 6\,400 & 0.432 & 0.926 & 1.140 & 0.280 &  0.082 \\
    & 6\,600 & 0.405 & 0.863 & 1.060 & 0.247 &  0.065 \\
    & 6\,800 & 0.380 & 0.804 & 0.984 & 0.218 &  0.049 \\
    & 7\,000 & 0.357 & 0.748 & 0.912 & 0.190 &  0.034 \\
    & 7\,200 & 0.337 & 0.694 & 0.845 & 0.164 &  0.019 \\
    & 7\,400 & 0.319 & 0.643 & 0.781 & 0.141 &  0.006 \\
    & 7\,600 & 0.303 & 0.593 & 0.720 & 0.118 & $-$0.007 \\
    & 7\,800 & 0.289 & 0.545 & 0.662 & 0.098 & $-$0.020 \\
    & 8\,000 & 0.276 & 0.499 & 0.606 & 0.078 & $-$0.032 \\
8.5 & 5\,800 & 0.542 & 1.128 & 1.409 & 0.396 &  0.136 \\
    & 6\,000 & 0.497 & 1.050 & 1.304 & 0.350 &  0.116 \\
    & 6\,200 & 0.461 & 0.982 & 1.215 & 0.311 &  0.098 \\
    & 6\,400 & 0.430 & 0.918 & 1.131 & 0.277 &  0.081 \\
    & 6\,600 & 0.403 & 0.857 & 1.053 & 0.245 &  0.065 \\
    & 6\,800 & 0.378 & 0.799 & 0.978 & 0.215 &  0.050 \\
    & 7\,000 & 0.355 & 0.744 & 0.908 & 0.188 &  0.035 \\
    & 7\,200 & 0.335 & 0.691 & 0.842 & 0.162 &  0.021 \\
    & 7\,400 & 0.316 & 0.641 & 0.779 & 0.139 &  0.008 \\
    & 7\,600 & 0.300 & 0.593 & 0.720 & 0.117 & $-$0.004 \\
    & 7\,800 & 0.286 & 0.546 & 0.663 & 0.097 & $-$0.016 \\
    & 8\,000 & 0.274 & 0.501 & 0.609 & 0.078 & $-$0.028 \\
\hline
\end{tabular}
\end{table}

\clearpage

\section{Abundance measurements}

\begin{table}
\caption{Abundance measurements of cool white dwarfs. \label{tbl-cool}}
\centering
\begin{tabular}{llcccc}
\hline\hline
WD & Name & $T_{\rm eff}$ (K) & $\log{\rm (Ca/H)}$ & $\log{\rm (Fe/H)}$ & Reference \\
\hline\\
0028$+$220     & NLTT1675     & 6010 & $-$9.53 & $-$8.63 & 1 \\
0032$-$175     & G266-135     & 9235 &$-$10.20 & \ldots  & 2 \\
0151$-$308     & NLTT6390     & 6020 &$-$10.00 & $-$8.57 & 1 \\
0208$+$396     & G74-7        & 7201 & $-$8.84 & $-$7.89 & 2 \\
0243$-$026     & LHS1442      & 6798 & $-$9.90 & \ldots  & 2 \\
0245$+$541     & G174-14      & 5190 &$-$11.70 & \ldots  & 2,3 \\
0315$-$293\tablefootmark{a} & NLTT10480 & 5200 &$-$10.30 & \ldots& 4 \\
0322$-$019\tablefootmark{a} & G77-50    & 5310 & $-$9.8  & $-$8.70 & 5 \\
0334$-$224     & NLTT11393    & 5880 &$-$10.24 & \ldots  & 1 \\
0543$+$579     & GD290        & 8142 &$-$10.29 & \ldots  & 2 \\
0846$+$346     & GD96         & 7373 & $-$9.41 & \ldots  & 2 \\
1054$-$226     & NLTT25792    & 8160 & $-$7.30 & \ldots  & 6 \\
1102$-$183     & EC11023-1821 & 8026 &$-$10.43 & \ldots  & 2 \\
1124$-$293     & ESO439-80    & 9680 & $-$8.53 & \ldots  & 2 \\
1202$-$232     & EC12028-2316 & 8619 & $-$9.78 & $-$8.00 & 2,6 \\
1208$+$576     & G197-47      & 5830 &$-$10.96 & \ldots  & 2 \\
1225$+$006     & HE1225+0038  & 9390 & $-$9.8  & \ldots  & 6 \\
1257$+$278     & G149-28      & 8600 & $-$8.04 & $-$7.41 & 7 \\
1315$-$110     & HE1315-1105  & 9080 & $-$9.4  & \ldots  & 6 \\
1344$+$106     & G63-54       & 6945 &$-$11.13 & \ldots  & 2 \\
1407$+$425     & PG           & 9856 & $-$9.87 & \ldots  & 2 \\
1455$+$298     & LHS3007      & 7366 & $-$9.31 & $-$8.60 & 2,6 \\
1633$+$433     & G180-63      & 6569 & $-$8.63 & $-$7.55 & 2 \\
1653$+$385\tablefootmark{a} & NLTT43806 & 5900 & $-$7.9  & $-$7.80 & 7 \\
1821$-$131     & LHS3384      & 7029 &$-$10.71 & \ldots  & 2 \\
1826$-$045     & G21-16       & 9480 & $-$8.83 & \ldots  & 2 \\
1858$+$393     & G205-52      & 9470 & $-$7.84 & $-$6.92 & 2 \\
2115$-$560     & LTT8452      & 9570 & $-$8.0  & $-$6.90 & 6 \\
2221$-$165     & HE2221-1630  & 9990 & $-$8.0  & \ldots  & 6 \\
\hline
\end{tabular}\\
\tablefoottext{a}{Magnetic}
\tablebib{ (1) This work; (2) \citet{zuc2003}; (3) \citet{ber2005}; 
(4) \citet{kaw2011a}; (5) \citet{far2011}; (6) \citet{kaw2011b}; 
(7) \citet{zuc2011}}
\end{table}

\end{document}